\documentclass[twocolumn,a4paper,pre,twoside,groupedaddress,showkeys,floatfix]{revtex4-1}

\usepackage{amsmath}
\usepackage{amssymb}
\usepackage{graphicx}
\usepackage{dcolumn}
\usepackage{bm}
\usepackage{xcolor}

\def\dps{\displaystyle}
\def\m#1{\mathrm{#1}}
\def\dd{\mathrm{d}}
\def\p{\partial}

\def\wh#1{\widehat{#1}}

\def\sign{\mathop\mathrm{sign}}

\def\arsinh{\mathop\mathrm{arsinh}}

\begin{document}


\title{Non-equilibrium steady states of electrolyte interfaces}

\author{Markus Bier}
\email{markus.bier@thws.de}
\affiliation
{
   Fakult\"{a}t Angewandte Natur- und Geisteswissenschaften,\\
   Technische Hochschule W\"{u}rzburg-Schweinfurt,\\
   Ignaz-Sch\"{o}n-Str.\ 11, 97421 Schweinfurt, Germany
}

\date{19 September 2023}

\begin{abstract}
The non-equilibrium steady states of a semi-infinite quasi-one-dimensional
univalent binary electrolyte solution, characterised by non-vanishing electric
currents, are investigated by means of Poisson-Nernst-Planck (PNP) theory.
Exact analytical expressions of the electric field, the charge density and
the number density are derived, which depend on the electric current density as
a parameter.
From a non-equilibrium version of the Grahame equation, which relates the total
space charge per cross-sectional area and the corresponding contribution of the
electric potential drop, the current-dependent differential capacitance of the
diffuse layer is derived.
In the limit of vanishing electric current these results reduce to those within
Gouy-Chapman theory.
It is shown that improperly chosen boundary conditions lead to non-equilibrium
steady state solutions of the PNP equations with negative ion number densities.
A necessary and sufficient criterion on surface conductivity constitutive
relations is formulated which allows one to detect such unphysical solutions.
\end{abstract}


\maketitle


\section{\label{sec:intro}Introduction}

The dynamics of ions in external electric fields determines the properties of
numerous important natural and technological processes such as the formation of
a membrane potential in biological cells via ion channels
\cite{Goldman1943, Arndt1970, Buck1976, Smith1977, Brumleve1978, Eisenberg1999,
Samin2015}, the charging and discharging of batteries by charge transfer
reactions at electrodes \cite{Vetter1967, Bagotsky2006}, the motion of colloids
exploiting electrokinetic effects of ionic solvent components \cite{Russel1989,
Hunter2001, Bazant2009a} as well as the suppression of electric currents by
insulation fluids in high electric fields \cite{Mie1908, Whitehead1930,
Felici1985, Lewis1994}.

Poisson-Nernst-Planck (PNP) theory \cite{Nernst1889, Planck1890, Russel1989} is
an established and widely used theoretical framework in order to address
these questions by considering the distributions of the electric field, the
charge density the ion number densities etc.
However, in studies of ion channels, high concentrated battery electrolytes and
colloid migration steric effects of ions have been noted to be relevant
\cite{Kornyshev2007, Kilic2007, Bazant2011} so that extensions of the PNP
theory are currently under investigation \cite{Gillespie2002, Garvish2018}.

In contrast, ideal insulation fluids for high voltage applications would be
void of ions and carefully prepared real insulation fluids contain only a few.
Hence ionic steric effects should be weak in that type of systems such that PNP
theory can be considered a well justified starting point \cite{Suh2012}.
However, in order to sustain the insulation property the ion concentration has
to stay low for long times.
Hence, understanding the mechanisms of charge generation in insulation fluids
is an important topic which has been under investigation for many decades
\cite{Yasufuku1979, Felici1985, Gafvert1992, Lewis1994, Castellanos1998,
Butcher2006, Sha2014}.

\begin{figure}
\includegraphics{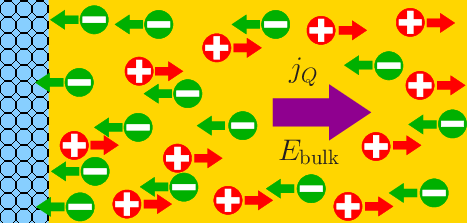}
\caption{A semi-infinite electrolyte solution of univalent cations (positive
ions, red) and univalent anions (negative ions, green) dispersed in a
dielectric continuum (solvent, yellow) is in contact with a planar electrode
(blue).
Charge transfer processes at the electrode surface and a non-vanishing electric
field $E_\text{bulk}$ in the bulk give rise to a non-equilibrium steady state
with a non-vanishing, spatially uniform charge current density $j_Q$ (violet
arrow).}
\label{fig:1}
\end{figure}

Unlike for nano-sized ion channels or colloids, the functioning of insulation
fluids requires a truly macroscopic spatial extension of length scales of the
order $1\,\m{cm}$ and above.
Hence a clear separation of length scales occurs: Within distances of molecular
size from (typically metallic) surfaces electrochemical processes can occur,
which provide the most important sources for ion generation, as electric fields
are strongest there.
Further away from the surfaces an extended bulk fluid with smooth distributions
of the electric field and ions is present.
The smoothness of these distributions allows for a description in terms of
(partial) differential equations and the smallness of the ion number densities
justifies the applicability of the PNP equations (see Sec.~\ref{sec:model:PNP}
below).
Moreover, the macroscopic character of high-voltage insulation systems quite
naturally motivates to study electric currents theoretically in terms of a
simple one-dimensional model comprising a planar electrode in contact with a
semi-infinite electrolyte solution, similar to the investigations of Gouy and
Chapman \cite{Gouy1909, Gouy1910, Chapman1913, Grahame1947, Russel1989,
Hunter2001} more than a century ago for conditions of thermodynamic equilibrium
(see Fig.~\ref{fig:1}).

Surprisingly, such a simple \emph{semi-infinite} Gouy-Chapman model for the
diffuse layer out of thermodynamic equilibrium has apparently \emph{not} yet
been studied analytically, whereas analytical solutions have well been obtained
for the mathematically more complicated situation of \emph{finite}
one-dimensional electrolytes, involving Jacobi elliptic functions, perturbation
expansions or simplifying approximations \cite{Malvadkar1972, Buck1973,
Leuchtag1977, Kosinska2008b, Golovnev2009, Kuzmin2010, Golovnev2010,
Golovnev2011, Shobukhov2014, Wang2014, Elad2019, Lyu2020, Asylamov2022}.

It is the purpose of the present work to analytically solve the PNP equations
(Sec.~\ref{sec:model:PNP}) for non-equilibrium steady states of a semi-infinite
quasi-one-dimensional univalent binary electrolyte solution
(Sec.~\ref{sec:model:setup}).
It turns out that the solutions can be represented in terms of elementary
functions and the expressions are of similar complexity as the Gouy-Chapman
results for thermodynamic equilibrium (Sec.~\ref{sec:model:solution}).
Moreover, a generalisation of the Grahame equation, which for thermodynamic
equilibrium expresses the surface charge density in terms of the surface
potential \cite{Grahame1947, Russel1989, Hunter2001}, to non-equilibrium
steady states is derived (Sec.~\ref{sec:model:Grahame}).
This allows to discuss the dependence of the distributions of the electric
field as well as of charge and ion number density on the electric current
(Sec.~\ref{sec:results:profiles}).
Moreover, it is shown that the profiles as well as the Grahame relation and
the differential capacitance of the space charge region reduce to the
well-known Gouy-Chapman results for vanishing electric current
(Sec.~\ref{sec:results:charge}).
A novel feature of the considered semi-infinite model, which is shown to occur
only out of equilibrium, is the occurrence of solutions of the PNP equations
with negative ion number densities (Sec.~\ref{sec:results:bc}).
This leads to the conclusion, that particular care is required in choosing
appropriate boundary conditions in order to avoid such unphysical solutions
(Sec.~\ref{sec:conclusions}).


\section{\label{sec:model}Model and formalism}

\subsection{\label{sec:model:setup}Setup}

Consider a semi-infinite univalent binary electrolyte solution, which is
bounded by a single planar electrode (see Fig.~\ref{fig:1}).
The solvent is described as a dielectric continuum of temperature $T$ and
dielectric constant $\varepsilon$.
Cations (positive ions, valency $Z_+=1$, diffusion constant $D_+$, number
density $\varrho_+$) and anions (negative ions, valency $Z_-=-1$, diffusion
constant $D_-$, number density $\varrho_-$) migrate in an electric field $E$,
which is oriented in normal direction of the electrode.
Due to the planar symmetry, the electric field $E(x)$ and the number densities
$\varrho_\pm(x)$ are functions of the distance $x\geq0$ from the electrode, but
they are independent of the lateral position.
Charge transfer processes close to the electrode surface allow for an electric
current to occur in the system.

It is assumed that a sufficiently long waiting time has elapsed such that the
system has attained a steady state in which the electric field $E$ and the
number densities $\varrho_\pm$ are time-independent.
Moreover, at large distances $x\to\infty$ from the electrode these quantities
are assumed to approach the constant bulk limits $E(x)\to E_\text{bulk}$ and
$\varrho_\pm(x)\to\varrho_\text{bulk}/2$.
Consequently, for $E_\text{bulk} = 0$ the system is in thermodynamic
equilibrium, whereas for $E_\text{bulk} \not= 0$ it is in a non-equilibrium
steady state.
In a non-equilibrium steady state a non-vanishing, spatially uniform electric
current density $j_Q\not=0$ is present, which requires concomitant charge
transfer processes to occur at the electrode surface.


\subsection{\label{sec:model:PNP}Governing equations}

In order to quantify the steady states of the system described in the previous
Sec.~\ref{sec:model:setup} the governing equations are derived within PNP
theory in the following.

Given ion number densities $\varrho_\pm(x)$ the \emph{total ion number density}
\begin{align}
   \varrho(x) = \varrho_+(x) + \varrho_-(x)
   \label{eq:rho}
\end{align}
and the \emph{charge density}
\begin{align}
   q(x) = e\big(\varrho_+(x) - \varrho_-(x)\big)
   \label{eq:q}
\end{align}
with the elementary charge $e$ are defined.
At large distances $x\to\infty$ from the electrode these quantities approach
the limits $\varrho(x)\to\varrho_\text{bulk}$ and $q(x)\to0$.

By means of Gauss' law \cite{Jackson1999} the derivative $E'(x)$ of the
electric field is related to the charge density $q(x)$:
\begin{align}
   \varepsilon_0\varepsilon E'(x) = q(x),
   \label{eq:Gauss}
\end{align}
where $\varepsilon_0$ is the vacuum permittivity.
This shows that approaching a constant bulk electric field $E(x)\to
E_\text{bulk}$ in the limit $x\to\infty$ implies local charge neutrality
$q(x) \to 0$ in the bulk.

The Nernst-Planck equations \cite{Nernst1889, Planck1890, Russel1989}
\begin{align}
   j_\pm(x) = D_\pm\big(-\varrho_\pm'(x) \pm \beta e \varrho_\pm(x) E(x)\big),
   \label{eq:NernstPlanck}
\end{align}
where $1/\beta = k_BT$ with the Boltzmann constant $k_B$ denotes the thermal
energy, describe the current densities of ions due to diffusion in a density
gradient and drift in the electric field.
Note that no advection contribution occurs in the present quiescent solvent.

In general the time dependence of the ion number densities $\varrho_\pm(x,t)$
is given by the continuity equations
\begin{align}
   \dot{\varrho}_\pm(x,t) = -j'_\pm(x,t),
   \label{eq:conti}
\end{align}
which describe conservation of the number of ions.
However, a steady state is time-independent ($\dot{\varrho}_\pm = 0$)
such that the current densities $j_\pm(x,t)$ are spatially uniform and
time-independent:
\begin{align}
   0 = j'_\pm(x,t)
   \qquad\Leftrightarrow\qquad
   j_\pm(x,t) = \text{const} = j_\pm.
   \label{eq:statjpm}
\end{align}
The same is true for the \emph{total number current density}
\begin{align}
   j_N = j_+ + j_-
   \label{eq:jN}
\end{align}
and the \emph{charge current density}
\begin{align}
   j_Q = e\big(j_+ - j_-\big).
   \label{eq:jQ}
\end{align}

In order to simplify the later calculations the following \emph{reduced
current densities} are introduced:
\begin{alignat}{2}
   J_N  &=\phantom{e\Big(}\frac{\dps j_+}{\dps D_+} + \frac{\dps j_-}{\dps D_-}
         \phantom{\Big)}
       &&= -\varrho'(x) + \phantom{e^2}\beta q(x)E(x)
   \label{eq:JN}\\
   J_Q  &= e\Big(\frac{\dps j_+}{\dps D_+} - \frac{\dps j_-}{\dps D_-}\Big)
       &&= -q'(x) + \beta e^2 \varrho(x)E(x).
   \label{eq:JQ}
\end{alignat}

As $\varrho'(x)\to0$ (due to $\varrho(x)\to\varrho_\text{bulk}$), $q(x)\to0$
and $E(x)\to E_\text{bulk}$ for $x\to\infty$, Eq.~\eqref{eq:JN} implies
$J_N = 0$, hence, from Eq.~\eqref{eq:JN},
\begin{align}
   \frac{j_+}{D_+} = -\frac{j_-}{D_-}.
   \label{eq:jpjm}
\end{align}
Consequently
\begin{align}
   j_Q = e\Big(1 + \frac{D_-}{D_+}\Big)j_+
       = -e\Big(\frac{D_+}{D_-} + 1\Big)j_-,
   \label{eq:jQfromjpm}
\end{align}
which allows one to express the ionic current densities $j_\pm$ in terms of the
charge current density $j_Q$.
This leads to the relation
\begin{align}
   J_Q = \frac{ej_+}{D_+} - \frac{ej_-}{D_-} = \frac{2}{D_+ + D_-} j_Q
       = \frac{j_Q}{D}
   \label{eq:JQjQ}
\end{align}
between the charge current density $j_Q$ and the reduced charge current density
$J_Q$ with the \emph{average diffusion constant}
\begin{align}
   D = \frac{D_+ + D_-}{2}.
   \label{eq:D}
\end{align}
   
Similarly, as $q'(x)\to0$ (due to $q(x)\to0$), $\varrho(x)\to
\varrho_\text{bulk}$ and $E(x)\to E_\text{bulk}$ for $x\to\infty$,
Eq.~\eqref{eq:JQ} leads to
\begin{align}
   J_Q = \beta e^2\varrho_\text{bulk}E_\text{bulk}.
   \label{eq:JQ-E}
\end{align}
By using Eq.~\eqref{eq:JQjQ} one infers
\begin{align}
   j_Q = D\beta e^2\varrho_\text{bulk}E_\text{bulk}
       = S_\text{bulk}E_\text{bulk}
\end{align}
with the \emph{bulk conductivity}
\begin{align}
   S_\text{bulk} = D\beta e^2\varrho_\text{bulk}.
   \label{eq:Sbulk}
\end{align}


\subsection{\label{sec:model:solution}Analytical solution}

In the following the PNP equations~\eqref{eq:Gauss}--\eqref{eq:conti} of the
considered system are solved analytically.

In a first step inserting Eq.~\eqref{eq:Gauss} in Eq.~\eqref{eq:JN} yields
\begin{align}
   0 &= -\varrho'(x) + \beta\varepsilon_0\varepsilon E'(x)E(x)
   \notag\\
     &= \Big(-\varrho(x) + \frac{\beta\varepsilon_0\varepsilon}{2}E(x)^2\Big)',
   \label{eq:dglrho}
\end{align}
which implies
\begin{align}
   -\varrho(x) + \frac{\beta\varepsilon_0\varepsilon}{2}E(x)^2
   = \text{const}.
   \label{eq:exprrho}
\end{align}
Evaluation of the constant by taking the limit $x\to\infty$ leads to
\begin{align}
   \varrho(x) =
   \varrho_\text{bulk} +
   \frac{\beta\varepsilon_0\varepsilon}{2}
   \big(E(x)^2 - E_\text{bulk}^2\big).
   \label{eq:rhobyE}
\end{align}

By inserting Eqs.~\eqref{eq:Gauss} and \eqref{eq:rhobyE} in
Eq.~\eqref{eq:JQ} one obtains the inhomogeneous non-linear ordinary
differential equation for the electric field $E(x)$
\begin{align}
   \varepsilon_0\varepsilon E''(x) =
   &\ \beta e^2\Big(\varrho_\text{bulk} -
       \frac{\beta\varepsilon_0\varepsilon}{2}E_\text{bulk}^2\Big)E(x)
   \notag\\
   &\ + \frac{\beta^2e^2\varepsilon_0\varepsilon}{2}E(x)^3 - J_Q.
   \label{eq:odeE}
\end{align}
This equation resembles Eq.~(A.9) of Ref.~\cite{Cohen1965} and Eq.~(48) of
Ref.~\cite{Buck1973}.

Introducing the excess electric field $\Delta E(x) = E(x) - E_\text{bulk}$
transforms the differential equation~\eqref{eq:odeE} to the homogeneous
differential equation
\begin{align}
   \Delta E''(x) =
   &\ \frac{\beta e^2}{\varepsilon_0\varepsilon}\big(\varrho_\text{bulk} +
      \beta\varepsilon_0\varepsilon E_\text{bulk}^2\big)\Delta E(x)
   \label{eq:odeDeltaE}\\
   &\ + \frac{3}{2}\beta^2e^2 E_\text{bulk}\Delta E(x)^2
      + \frac{1}{2}\beta^2e^2 \Delta E(x)^3.
   \notag
\end{align}

With the Debye length $1/\kappa$ defined by $\kappa^2 = 4\pi\ell_B
\varrho_\text{bulk}$, where $\ell_B = \beta e^2/(4\pi\varepsilon_0\varepsilon)$
is the Bjerrum length \cite{Debye1923, Russel1989, McQuarrie2000, Hunter2001},
one can introduce the dimensionless \emph{electric flux} parameter
\begin{align}
   \eta = \frac{\beta e j_Q}{\kappa S_\text{bulk}}
        = \frac{\beta e E_\text{bulk}}{\kappa},
   \label{eq:eta}
\end{align}
which quantifies the deviation of a steady state from thermodynamic equilibrium
($\eta=0$) in terms of the charge current density $j_Q$.
The values $\eta = \pm1$ correspond to a charge current density $j_Q =
\pm S_\text{bulk}\kappa/(\beta e)$ generated by a bulk electric field
$E_\text{bulk} = \pm\kappa/(\beta e)$.
The notion of electric flux $\eta$ allows one to rewrite
Eq.~\eqref{eq:odeDeltaE} in the form
\begin{align}
   \Delta E''(x) =
   &\ \kappa^2(1+\eta^2)\Delta E(x)
   \label{eq:odeDeltaEeta}\\
   &\ + \frac{3}{2}\beta e\kappa\eta\Delta E(x)^2
      + \frac{1}{2}\beta^2e^2 \Delta E(x)^3.
   \notag
\end{align}

Finally the transformation
\begin{align}
   y = \kappa x\sqrt{1+\eta^2},
   \qquad
   \wh{E}(y) = \frac{\beta e \Delta E(x)}{\kappa}
   \label{eq:trafo}
\end{align}
is used to rewrite Eq.~\eqref{eq:odeDeltaEeta} in the form
\begin{align}
   \wh{E}''(y) = \wh{E}(y) + \frac{3\eta}{2(1+\eta^2)}\wh{E}(y)^2
                 + \frac{1}{2(1+\eta^2)}\wh{E}(y)^3.
   \label{eq:odeEhat}
\end{align}

A first integral of Eq.~\eqref{eq:odeEhat} is found by multiplication with
$\wh{E}'(y)$ and the integration constant is fixed by using $\wh{E}'(y)\to0$
and $\wh{E}(y)\to0$ for $y\to\infty$:
\begin{align}
   \wh{E}'(y)^2 = \wh{E}(y)^2
   \Big(1+\frac{\eta}{1+\eta^2}\wh{E}(y) +
        \frac{1}{4(1+\eta^2)}\wh{E}(y)^2\Big).
   \label{eq:firstintegral}
\end{align}
It can be shown, that the expression inside the parentheses is bounded from
below by $1/(1+\eta^2)$ for any value of $\wh{E}(y)$, hence
\begin{align}
   |\wh{E}'(y)| \geq |\wh{E}(y)|/\sqrt{1+\eta^2}
   \qquad\text{for all $y\geq0$.}
   \label{eq:magnitudes}
\end{align}
Moreover, from Eq.~\eqref{eq:odeEhat} one infers that for $y\to\infty$ the
asymptotic behaviour of $\wh{E}(y)$ is monotonic, i.e.\ constant or exponential,
but not oscillatory.
Consequently, due to the limit $\wh{E}(y)\to0$ for $y\to\infty$, only three
cases can occur: $\wh{E}(y)$ for $y\geq0$ is either (i) constantly zero
($\wh{E}(y)=0$) or (ii) positive ($\wh{E}(y)>0$) and strictly monotonically
decreasing ($\wh{E}'(y)<0$) or (iii) negative ($\wh{E}(y)<0$) and strictly
monotonically increasing ($\wh{E}'(y)>0$).
Otherwise if, say, $\wh{E}(y)$ approaches the limit $\wh{E}(y)\to0$ for
$y\to\infty$ from above, but $\wh{E}(y)$ was not monotonically decreasing for
all $y\geq0$, there would be a local maximum at some position $y=y^*$ with
$\wh{E}'(y^*)=0$ but $\wh{E}(y^*)>0$, which contradicts
Eq.~\eqref{eq:magnitudes}.
A similar contradiction arises from the assumption of a non-monotonically
increasing behaviour when approaching the bulk limit from below.

In the above cases (ii) and (iii) $\wh{E}'(y)$ and $\wh{E}(y)$ have opposite
sign, so that Eq.~\eqref{eq:firstintegral} leads to
\begin{align}
   \wh{E}'(y) = - \wh{E}(y)\sqrt{1+\frac{\eta}{1+\eta^2}\wh{E}(y) +
                                 \frac{1}{4(1+\eta^2)}\wh{E}(y)^2},
   \label{eq:EdivE}
\end{align}
which relates the value $\wh{E}(y)$ and the derivative $\wh{E}'(y)$.
This equation is obviously true also for case (i).

For cases (ii) and (iii) separation of variables leads to the solution
\cite{Gradshteyn1980}
\begin{widetext}
\begin{align}
   \wh{E}(y) = \frac{\dps 2(1+\eta^2)\wh{E}_0}
   {\dps \sinh(y)\sqrt{\wh{E}_0^2+\big(2(1+\eta^2)+\wh{E}_0\eta\big)^2}
         + \cosh(y)\big(2(1+\eta^2)+\wh{E}_0\eta\big) - \wh{E}_0\eta}
   \label{eq:Ehat}
\end{align}
\end{widetext}
with the value $\wh{E}(0)=\wh{E}_0$ at the electrode surface $y=0$ playing the
role of the integration constant.
The solution $\wh{E}(y)\equiv0$ of case (i) is obtained from Eq.~\eqref{eq:Ehat}
with $\wh{E}_0 = 0$.


\subsection{\label{sec:model:Grahame}Grahame equation}

An important result in the thermodynamic equilibrium theory of electrolyte
interfaces is an expression for the surface charge density in terms of the
surface potential, which is commonly referred to as \emph{Grahame equation}
\cite{Grahame1947, Russel1989, Hunter2001}.
Here an analogous expression is derived for steady state conditions.

Integration of the excess electric field $\Delta E(x)$ leads to the
\emph{excess voltage}
\begin{align}
   \Delta U = \int\limits_0^\infty \dd x\,\Delta E(x),
   \label{eq:DeltaU}
\end{align}
which measures deviations from local charge neutrality expressed by the charge
density $q(x) = \varepsilon_0\varepsilon\Delta E'(x)$.
It adds to the voltage required to sustain the bulk electric field
$E_\text{bulk}$.
Using Eqs.~\eqref{eq:trafo} and \eqref{eq:EdivE} one obtains
\cite{Gradshteyn1980}
\begin{align}
   \beta e \Delta U
   &=
   \frac{\dps 1}{\dps \sqrt{1+\eta^2}}
   \int\limits_0^\infty \dd x\,\kappa\sqrt{1+\eta^2}\
                              \frac{\dps \beta e \Delta E(x)}{\dps \kappa}
   \notag\\
   &=
   \frac{\dps 1}{\dps \sqrt{1+\eta^2}}
   \int\limits_0^\infty \dd y\,\wh{E}(y)
   \notag\\
   &=
   -\frac{\dps 1}{\dps \sqrt{1+\eta^2}}
   \int\limits_0^\infty \dd y\,\frac{\dps\wh{E}'(y)}
   {\dps\sqrt{1 + \frac{\dps\eta\wh{E}(y)}{\dps 1+\eta^2} +
              \frac{\dps\wh{E}(y)^2}{\dps 4(1+\eta^2)}}}
   \notag\\
   &=
   2\Big(\arsinh\Big(\frac{1}{2}\wh{E}(0)+\eta\Big) - \arsinh(\eta)\Big).
   \label{eq:DeltaUEhat0}
\end{align}

Using Eq.~\eqref{eq:Gauss} the total charge per cross-sectional area of the
diffuse layer can be calculated by
\begin{align}
   \int\limits_0^\infty\dd x\,q(x) =
   \varepsilon_0\varepsilon\int\limits_0^\infty\dd x\,\Delta E'(x) =
   -\varepsilon_0\varepsilon\Delta E(0).
   \label{eq:totalcharge}
\end{align}
This total charge per cross-sectional area of the space charge region is
balanced by an \emph{excess surface charge density} $\Delta\sigma$ at the
electrode surface, which has the same magnitude but the opposite sign:
$\Delta\sigma = \varepsilon_0\varepsilon\Delta E(0)$.
It expresses the excess of the \emph{total surface charge density}
\begin{align}
   \sigma = \sigma_\text{bulk} + \Delta\sigma
   \label{eq:sigma}
\end{align}
compared to the surface charge density $\sigma_\text{bulk} = \varepsilon_0
\varepsilon E_\text{bulk}$, which generates the bulk electric field
$E_\text{bulk}$.
By introducing the \emph{saturation charge density} \cite{Bocquet2002}
\begin{align}
   \sigma_\text{sat} = \frac{e\kappa}{\pi\ell_B}
                     = 4\varepsilon_0\varepsilon\frac{\kappa}{\beta e}
   \label{eq:sigmasat}
\end{align}
one obtains the relation
\begin{align}
   \wh{E}(0) = \frac{\beta e\Delta E(0)}{\kappa}
             = 4 \frac{\dps\Delta\sigma}{\dps \sigma_\text{sat}}
   \label{eq:EhatDeltasigma}
\end{align}
by means of which Eq.~\eqref{eq:DeltaUEhat0} can be rewritten as
\begin{align}
   \beta e \Delta U =
   2\Big(\arsinh\Big(2\frac{\dps\Delta\sigma}{\dps\sigma_\text{sat}}+\eta\Big)
         - \arsinh(\eta)\Big).
   \label{eq:DeltaUDeltasigma}
\end{align}

Solving Eq.~\eqref{eq:DeltaUDeltasigma} for the excess surface charge density
$\Delta\sigma$ one obtains the \emph{Grahame equation} 
\begin{align}
   \Delta\sigma =
   \frac{\dps\sigma_\text{sat}}{\dps 2}\Big(
   &\sqrt{1+\eta^2}\sinh\Big(\frac{\beta e \Delta U}{2}\Big)
   \notag\\
   &+ \eta\Big(\cosh\Big(\frac{\beta e \Delta U}{2}\Big) - 1\Big)\Big).
   \label{eq:Grahame}
\end{align}

The derivative of the excess surface charge density $\Delta\sigma$ w.r.t.\ the
excess voltage $\Delta U$ leads to the \emph{differential capacitance} of the
diffuse layer
\begin{align}
   C
   &= \frac{\p\Delta\sigma}{\p\Delta U}
   \label{eq:capacitance}\\
   &= \frac{\dps\beta e \sigma_\text{sat}}{\dps 4}
      \Big(\!\sqrt{1+\eta^2}\cosh\Big(\frac{\beta e \Delta U}{2}\Big)
           \!+\! \eta\sinh\Big(\frac{\beta e \Delta U}{2}\Big)\Big).
   \notag
\end{align}


\section{\label{sec:results}Results and Discussion}

\subsection{\label{sec:results:profiles}Steady state distributions}

\begin{figure}
\includegraphics{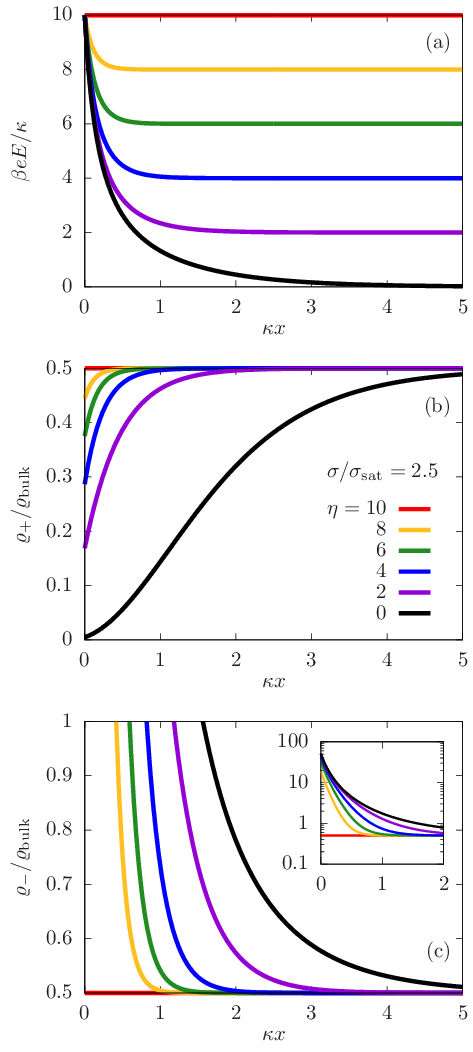}
\caption{Steady state distributions of (a) the electric field $E$, (b) the
cation number density $\varrho_+$ and (c) the anion number density $\varrho_-$
as functions of the normal distance $x$ from the electrode surface for fixed
total surface charge density $\sigma = 2.5\,\sigma_\text{sat}$ and various
values of the electric flux $\eta$ defined in Eq.~\eqref{eq:eta}.
These quantities approach their bulk values on a length scale
$\lambda(\eta)$, which decreases with increasing magnitude $|\eta|$ of the bulk
flux (see Fig.~\ref{fig:3}).
For $E_\text{bulk}=0$ (thermodynamic equilibrium) the electric field $E(x)$
within Gouy-Chapman theory is reproduced, whereas purely exponential solutions
are approached for $\eta \gg 1$.}
\label{fig:2}
\end{figure}

Some typical examples of steady state distributions of the electric field
$E(x)$, the cation number density $\varrho_+(x) = (\varrho(x)+q(x)/e)/2$ and
the anion number density $\varrho_-(x) = (\varrho(x)-q(x)/e)/2$ are displayed
in Fig.~\ref{fig:2}.
These profiles correspond to the largely arbitrary choice of total surface
charge density $\sigma = 2.5\,\sigma_\text{sat}$ and electric flux $\eta\in\{0,
2,4,6,8,10\}$.

Hypothetical pure water of $p\text{H}=7$, i.e.\ with total ion number density
in the bulk of $\varrho_\text{bulk} = 2\cdot10^{-7}\,\m{M}$, at $T=300\,\m{K}$
has Bjerrum length $\ell_B\approx0.7\,\m{nm}$ and Debye length $1/\kappa\approx
1\,\m{\mu m}$ so that $\sigma_\text{sat}\approx7\cdot10^{-5}\,\m{C/m^2}$, which
corresponds to an electric field $E_\text{sat}=\sigma_\text{sat}/(\varepsilon_0
\varepsilon)\approx 100\,\m{V/mm}$ at the electrode.
Furthermore, assuming a bulk conductivity $S_\text{bulk}=5\cdot10^{-6}\,\m{S/m}$
(see Ref.~\cite{Lide1998}) an electric flux $\eta=1$ corresponds to a charge
current density of $j_Q \approx 0.1\,\m{A/m^2}$.
However, these values can vary in a wide range for various materials, and the
purpose of the present work is not to discuss a particular system in detail.

The electric field $E(x) = E_\text{bulk} + \Delta E(x)$ is obtained from the
analytical solution Eq.~\eqref{eq:Ehat} in conjunction with
Eq.~\eqref{eq:trafo}.
The total number density $\varrho(x)$ is obtained from Eq.~\eqref{eq:rhobyE}
and the charge density $q(x)$ is calculated via Eq.~\eqref{eq:Gauss}. 

For $\eta=0$ (thermodynamic equilibrium, $j_Q=0$) the bulk electric field
vanishes, $E_\text{bulk}=0$ (see Eq.~\eqref{eq:eta}), hence $\Delta E(x) =
E(x)$, and Eq.~\eqref{eq:Ehat} reduces to (see Eq.~\eqref{eq:trafo})
\begin{align}
   \frac{\beta e \Delta E(x)}{\kappa}
   = \frac{\dps\wh{E}_0}{\dps\sqrt{1+\Big(\frac{1}{2}\wh{E}_0\Big)^2}
     \sinh(\kappa x) + \cosh(\kappa x)},
   \label{eq:Eeq}
\end{align}
which coincides with the electric field derived within Gouy-Chapman theory
\cite{Gouy1909, Gouy1910, Chapman1913, Grahame1947, Russel1989, Hunter2001}.

For $|\eta|\gg1$ Eq.~\eqref{eq:Ehat} simplifies to
\begin{align}
   \frac{\beta e \Delta E(x)}{\kappa}
   \simeq \wh{E}_0\exp\Big(-\frac{x}{\lambda(\eta)}\Big)
\end{align}
with the length scale
\begin{align}
   \lambda(\eta) = \frac{1}{\kappa\sqrt{1+\eta^2}},
   \label{eq:lambda}
\end{align}
i.e.\ the excess electric field $\Delta E(x)$ becomes purely exponential.
This is consistent with the fact that Eq.~\eqref{eq:odeEhat} reduces to the
linear equation $\wh{E}''(y) = \wh{E}(y)$ for $|\eta|\to\infty$.
It is remarkable that the governing equation~\eqref{eq:odeEhat} becomes simple
far away from thermodynamic equilibrium.

According to Eq.~\eqref{eq:trafo} the position dependence of the excess
electric field $\Delta E(x)$ for arbitrary values of the  electric flux $\eta$
is determined by length scale $\lambda(\eta)$ in Eq.~\eqref{eq:lambda}:
\begin{align}
   \frac{\beta e \Delta E(x)}{\kappa}
   = \wh{E}\Big(\frac{x}{\lambda(\eta)}\Big).
   \label{eq:scalingrelation}
\end{align}
In particular, according to Eq.~\eqref{eq:Ehat}, for $x \gg \lambda(\eta)$,
$\Delta E(x)$ exhibits an exponential asymptotic decay on the length scale
$\lambda(\eta)$.

\begin{figure}
\includegraphics{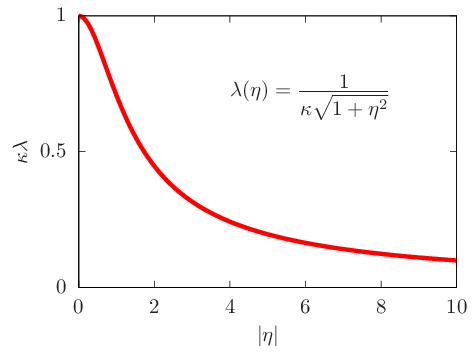}
\caption{Decay length $\lambda$ of correlations in a steady state with
electric flux parameter of magnitude $|\eta|$ (see Eq.~\eqref{eq:eta}).
For $\eta=0$ (thermodynamic equilibrium) it is identical to the Debye length
$1/\kappa$.}
\label{fig:3}
\end{figure}

The dependence of length scale $\lambda(\eta)$ on the electric flux parameter
$\eta$ is displayed in Fig.~\ref{fig:3}.
For $\eta=0$ (thermodynamic equilibrium, $j_Q=0$) this equals the Debye length
$\lambda(0) = 1/\kappa$, whereas it decreases $\lambda(\eta) \simeq
1/(\kappa|\eta|)$ for $|\eta|\to\infty$.
However, the detailed position dependence of $\Delta E(x)$ is irrelevant from
the practical point of view if $|\eta|$ is so large that the length scale
$\lambda(\eta)$ is of molecular size or below. 

In summary, with increasing electric flux $|\eta|\sim|j_Q|$ the electric field
changes from the Gouy-Chapman form in thermodynamic equilibrium to a purely
exponential dependence far away from thermodynamic equilibrium.
In parallel the corresponding relevant length scale $\lambda(\eta)$ decays from
the Debye length $1/\kappa$ to zero.


\subsection{\label{sec:results:charge}Space charge}

At distances $x \gg \lambda(\eta)$ the electric field and the densities become
spatially uniform (see Fig.~\ref{fig:2}).
Depending on the processes taking place at the electrode surface deviations
from spatial uniformity can occur there, which are commonly referred to as the
formation of \emph{space charges}.
In the theory of electrolyte solutions in thermodynamic equilibrium this space
charge region is traditionally called the \emph{diffuse layer}
\cite{Stern1924, Grahame1947, Russel1989, Hunter2001}.

In the notation of Sec.~\ref{sec:model:Grahame} the amount of space charge per
cross-sectional area is given by $-\Delta\sigma$, where $\Delta\sigma$ denotes
the excess surface charge density, which is the charge per cross-sectional area
on the electrode surface induced by the space charge.
The relation of $\Delta\sigma$ to the excess voltage $\Delta U$ of the space
charge region (see Eq.~\eqref{eq:DeltaU}) is called the Grahame
equation~\eqref{eq:Grahame}, in analogy to the equation of the same name within
the theory of electrolyte solutions in thermodynamic equilibrium
\cite{Grahame1947, Russel1989, Hunter2001}.

\begin{figure}
\includegraphics{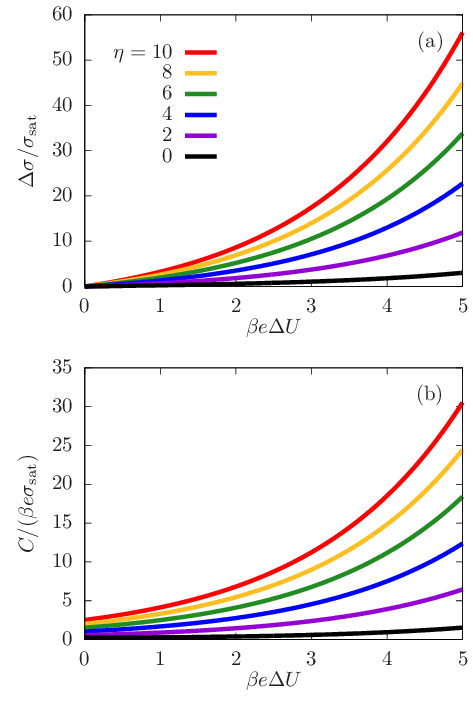}
\caption{(a) Excess surface charge density $\Delta\sigma$ and (b) differential
capacitance of the diffuse space charge region close to the electrode as
functions of the corresponding voltage drop $\Delta U$ for various values of
the electric flux $\eta$ in Eq.~\eqref{eq:eta}.
For $\eta=0$ (thermodynamic equilibrium) these quantities are identical to
those within Gouy-Chapman theory.}
\label{fig:4}
\end{figure}

Figure~\ref{fig:4}(a) displays this relation for electric flux $\eta\in\{0,2,4,
6,8,10\}$.
For $\eta=0$ (thermodynamic equilibrium, $j_Q=0$) the traditional Grahame
equation is found, whereas for a given excess voltage $\Delta U>0$ the amount of
space charge increases upon increasing the electric flux $\eta$.
According to Eq.~\eqref{eq:Grahame}, for $\eta\to\infty$ this increase is
asymptotically proportional to $\eta$.
Hence the mean capacitance $\Delta\sigma/\Delta U$ increases upon increasing
the electric flux, which can be attributed to the decrease of the diffuse layer
thickness $\lambda(\eta)$ upon increasing $|\eta|$.

Figure~\ref{fig:4}(b) displays the differential capacitance $C$, i.e.\ the
derivative of the excess surface charge $\Delta\sigma$ w.r.t.\ the excess
voltage $\Delta U$, given in Eq.~\eqref{eq:capacitance}.
For $\eta=0$ (thermodynamic equilibrium, $j_Q=0$) the well-known
Gouy-Chapman capacitance is found \cite{Russel1989, Hunter2001}, whereas $C$
increases upon increasing the electric flux $\eta$.
From Eq.~\eqref{eq:capacitance} one infers an asymptotically proportional
dependence of $C$ on $\eta$ for $\eta\to\infty$.
This dependence on $\eta$ can again be attributed to the decrease of the
diffuse layer thickness $\lambda(\eta)$ upon increasing $|\eta|$.

Note that within PNP theory no packing effects due to finite molecular sizes
of the ions are taken into account.
This precludes the decrease of the differential capacitance $C$ for large
values of the excess voltage $\Delta U$, which otherwise would set in once the
inner Helmholtz plane is fully occupied such that additionally adsorbed ions
have to be accommodated at larger distances from the electrode surface
\cite{Kornyshev2007, Fedorov2008a, Fedorov2008b, Bazant2011}.


\subsection{\label{sec:results:bc}Surface conductivity models}

Physically meaningful number densities $\varrho_\pm(x)$ have to be
non-negative, i.e.\ $\varrho_\pm(x)\geq0$, everywhere.
Using Eqs.~\eqref{eq:rho} and \eqref{eq:q} to rewrite this condition as
$\varrho_\pm = (\varrho \pm q/e)/2 \geq 0$, leads to $e\varrho \geq \mp q$,
which is equivalent to $e\varrho
\geq |q|$.
The latter inequality in turn is equivalent to the two conditions
\begin{align}
   \varrho \geq 0 \qquad\text{and}\qquad e|\varrho| \geq |q|.
   \label{eq:ineq}
\end{align}
Writing
\begin{align}
   \varrho &= \varrho_\text{bulk}\Big(1 + \eta\wh{E} + \frac{1}{2}\wh{E}^2\Big)
   \label{eq:rhobyEhat}\\   
   q       &= -e\varrho_\text{bulk}\wh{E}
           \sqrt{1+\eta^2+\eta\wh{E}+\frac{1}{4}\wh{E}^2}
   \label{eq:qbyEhat}
\end{align}
by using Eqs.~\eqref{eq:rhobyE}, \eqref{eq:Gauss} and \eqref{eq:EdivE} one
can reformulate $e|\varrho|\geq|q|$, i.e.\ $e^2\varrho^2 \geq q^2$, as
\begin{align}
   1 + 2\eta\wh{E} \geq 0.
   \label{eq:physokE}
\end{align}
Moreover, if Eq.~\eqref{eq:physokE}, i.e.\ $e|\varrho|\geq|q|$, is fulfilled,
one immediately finds from Eq.~\eqref{eq:rhobyEhat} and $\wh{E}^2 \geq 0$ that
\begin{align}
   \varrho =
   \frac{\varrho_\text{bulk}}{2}(1 + \underbrace{1 + 2\eta\wh{E}}_{\geq 0}
   + \wh{E}^2) \geq 0,
   \label{eq:rhononneg}
\end{align}
i.e.\ the second inequality in Eq.~\eqref{eq:ineq} implies the first.
To summarise the above reasoning: The physical condition of non-negative number
densities $\varrho_\pm$ is fulfilled if and only if Eq.~\eqref{eq:physokE}
holds.

Whereas Eq.~\eqref{eq:physokE} is fulfilled for $\eta\wh{E}\geq0$, it might be
violated for $\eta\wh{E}<0$.
As $\wh{E}$ is monotonic (see Sec.~\ref{sec:model:solution})
Eq.~\eqref{eq:physokE} is fulfilled if and only if it holds at the electrode
surface, i.e.\ $1+2\eta\wh{E}(0)\geq0$.
Using Eq.~\eqref{eq:EhatDeltasigma} the physical requirement
Eq.~\eqref{eq:physokE} of non-negative number densities $\varrho_\pm$ can be
formulated in terms of the excess surface charge density:
\begin{align}
   1 + 8\eta\frac{\Delta\sigma}{\sigma_\text{sat}}\geq 0.
   \label{eq:physoksigma}
\end{align}

\begin{figure}
\includegraphics{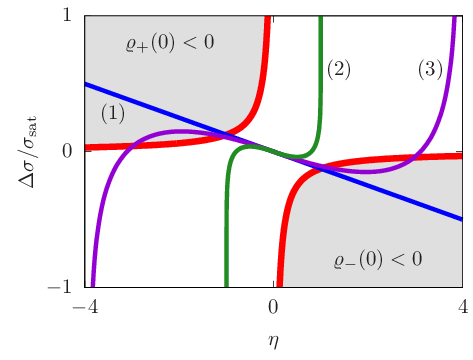}
\caption{Relations between excess surface charge density $\Delta\sigma$ and
electric flux $\eta$ (see Eq.~\eqref{eq:eta}) for three cases of surface
conductivity models:
(1) linear relation with surface-to-bulk conductivity ratio
$S_\text{surf}/S_\text{bulk} = 2$ (blue line),
(2) diffusion limited process with small saturation current (green curve) and
(3) diffusion limited process with large saturation current (violet curve).
The grey regions, bounded by red curves, correspond to unphysical conditions
where solutions of the PNP equations~\eqref{eq:Gauss}--\eqref{eq:conti} occur
which exhibit negative values of the number densities $\varrho_\pm(0)$ close to
the electrode.}
\label{fig:5}
\end{figure}

Figure~\ref{fig:5} provides a graphical representation of
Eq.~\eqref{eq:physoksigma} in the $\Delta\sigma$-$\eta$-plane.
For $\eta\not=0$ (non-equilibrium steady state, $j_Q\not=0$) condition
Eq.~\eqref{eq:physoksigma} is equivalent to
\begin{align}
   \frac{\Delta\sigma}{\sigma_\text{sat}} &\geq -\frac{1}{8\eta}
   \qquad\text{for $\eta>0$ and}
   \label{eq:physokpos}\\
   \frac{\Delta\sigma}{\sigma_\text{sat}} &\leq -\frac{1}{8\eta}
   \qquad\text{for $\eta<0$.}
   \label{eq:physokneg}
\end{align}
If Eq.~\eqref{eq:physokpos} is violated the PNP solution leads to
$\varrho_-(0) < 0$, whereas if Eq.~\eqref{eq:physokneg} is violated the PNP
solution yields $\varrho_+(0) < 0$ (see the grey regions in Fig.~\ref{fig:5}).
Note that the parameter range $\eta,\Delta\sigma,\Delta U\geq0$ of the examples
in Secs.~\ref{sec:results:profiles} and \ref{sec:results:charge} has been
chosen on purpose in order to fulfil condition Eq.~\eqref{eq:physoksigma}.

For $\eta=0$ (thermodynamic equilibrium, $j_Q=0$) condition
Eq.~\eqref{eq:physoksigma} is fulfilled for all excess surface charge densities
$\Delta\sigma$, i.e.\ negative number density solutions within PNP theory can
occur for non-equilibrium steady states, but not for states in thermodynamic
equilibrium.

In order to demonstrate the applicability of the previous result consider the
particular case of surface processes which give rise to a strictly linear
constitutive relation between charge current $j_Q$ and total electric field
$E(0)$ at the electrode surface:
\begin{align}
   j_Q = S_\text{surf} E(0).
   \label{eq:linsurfcondjE}
\end{align}
The proportionality factor $S_\text{surf}$ is called the \emph{surface
conductivity} here.
Using Eqs.~\eqref{eq:eta} and \eqref{eq:EhatDeltasigma} one can rewrite
Eq.~\eqref{eq:linsurfcondjE} of the linear surface conductivity model as
\begin{align}
   \frac{\Delta\sigma}{\sigma_\text{sat}}
   = \frac{1}{4}\Big(\frac{S_\text{bulk}}{S_\text{surf}} - 1\Big)\eta.
   \label{eq:linsurfcond}
\end{align}
The case of a surface-to-bulk conductivity ratio  $S_\text{surf}/S_\text{bulk}
= 2$ is represented by the blue line labelled with ``(1)'' in Fig.~\ref{fig:5}.
The fact that this line crosses over into the grey regions, where the
conditions Eqs.~\eqref{eq:physokpos} and \eqref{eq:physokneg} are violated,
shows that the purely linear surface conductivity model cannot be applied under
these conditions for too large electric fluxes.
Obviously the same argument applies to any system with $S_\text{surf} >
S_\text{bulk}$ (high surface conductivity), because then the slope of the
corresponding line is negative so that intersections with the unphysical grey
regions occur for sufficiently large electric flux $|\eta|$.
It should be noted that in the grey regions of Fig.~\ref{fig:5} no mathematical
problems arise: Equations~\eqref{eq:Ehat}, \eqref{eq:rhobyEhat} and
\eqref{eq:qbyEhat} are the solutions of the PNP equations for steady states
(see Sec.~\ref{sec:model:PNP}).
But these steady state solutions of the PNP equations may be physically
meaningless due to the occurrence of negative number densities.

In systems with $S_\text{surf} \leq S_\text{bulk}$ (low surface conductivity)
the linear model Eq.~\eqref{eq:linsurfcondjE} leads to a line
Eq.~\eqref{eq:linsurfcond} in Fig.~\ref{fig:5} with non-negative slope, which
does not intersect the grey regions, i.e.\ no negative number densities occur
under these conditions.
However, it is possible that other quantities exist, for which the PNP
solutions exhibit unphysical properties.
Moreover, whether the linear surface conductivity model
Eq.~\eqref{eq:linsurfcondjE}, even if no unphysical values occur, is able to
quantitatively describe real systems is an unrelated question.

In fact, the linear surface conductivity model Eq.~\eqref{eq:linsurfcondjE} can
be expected to be an acceptable description for sufficiently small surface
fields only, because for large surface fields saturation of the charge density
current $j_Q$ sets in due to an exhaustion of ions.
Such diffusion limited surface processes can be described by the constitutive
relation \cite{Vetter1967, Bagotsky2006}
\begin{align}
   E(0) = -\sign(j_Q)\frac{\dps j_{Q\text{sat}}}{\dps S_\text{surf}}
          \ln\Big(1-\frac{\dps |j_Q|}{\dps j_{Q\text{sat}}}\Big),
   \label{eq:diflimjE}
\end{align}
where $j_{Q\text{sat}}>0$ is the saturation charge current density and
$S_\text{surf}$ is the differential conductivity in the limit of infinitesimal
surface electric fields.
In Fig.~\ref{fig:5} the case of $S_\text{surf}/S_\text{bulk}=2$ for two
different values of the saturation charge current density $j_{Q\text{sat}}$ are
shown by the curves labelled ``(2)'' and ``(3)''.
The green curve ``(2)'' demonstrates that for sufficiently small
$|j_{Q\text{sat}}|$ no negative ion number densities occur, although a highly
conductive surface is present at weak surface fields.
However, the violet curve ``(3)'' for too large $|j_{Q\text{sat}}|$ does
exhibit unphysical solutions inside the grey regions.
Hence, great care is required to choose appropriate surface conductivity models,
which, in conjunction with the PNP equations, lead to physical solutions.


\section{\label{sec:conclusions}Conclusions}

In the present work the analytical solution Eq.~\eqref{eq:Ehat} of the PNP
equations for steady states (see Sec.~\ref{sec:model:PNP}) of a semi-infinite
univalent binary electrolyte solution in contact with a planar electrode
(Fig.~\ref{fig:1}) has been derived.
It can be expected to play a similar role as the Gouy-Chapman solution
\cite{Gouy1909, Gouy1910, Chapman1913, Grahame1947, Russel1989, Hunter2001}
for thermodynamic equilibrium, to which the derived solutions reduce for the
case of a vanishing charge current density (Fig.~\ref{fig:2}).
The characteristic length scale of the electric field as well as the number and
charge density profiles, which in thermodynamic equilibrium is given by
the Debye length, decreases for non-equilibrium steady states upon increasing
the magnitude of the charge current density (Fig.~\ref{fig:3}).
The Grahame equation, which expresses the surface charge density at the
electrode in terms of the voltage \cite{Grahame1947, Russel1989, Hunter2001},
is generalised to non-equilibrium steady states (Eq.~\eqref{eq:Grahame}).
The excess surface charge density at the electrode and the differential
capacitance of the space charge layer for given excess voltage are found to
vary with the current charge density of non-equilibrium steady states
(Fig.~\ref{fig:4}).
Finally it is found that, in contrast to the case of thermodynamic equilibrium
within Gouy-Chapman theory \cite{Gouy1909, Gouy1910, Chapman1913, Grahame1947,
Russel1989, Hunter2001}, the excess surface charge density may not take an
arbitrary value for a given non-vanishing charge current density of a
non-equilibrium steady state: Steady state solutions of the PNP equations
exist which give rise to physically meaningless negative ion number densities
(Fig.~\ref{fig:5}).
A concise criterion is formulated which can serve to identify such unphysical
solutions (Eq.~\eqref{eq:physoksigma}).

The most remarkable observation of the present work, i.e.\ the existence of
steady state solutions of the PNP equations which are physically meaningless,
calls for further investigation.
Two approaches are conceivable:
First, as Gauss' law Eq.~\eqref{eq:Gauss} and the continuity
equation~\eqref{eq:conti} are unexceptionable for fundamental physical reasons,
one could suggest to modify the Nernst-Planck equation~\eqref{eq:NernstPlanck}
in order to avoid unphysical solutions.
Second, one could try to keep the Nernst-Planck equation~\eqref{eq:NernstPlanck}
unchanged, but require boundary conditions to fulfil
Eq.~\eqref{eq:physoksigma}.
Whereas the second approach is the more pragmatic one, it will be the first one
which provides more fundamental insight into the non-equilibrium properties of
electrolyte solutions.



\begin{thebibliography}{00}

\bibitem{Goldman1943}
   D.E.\ Goldman,
   \textit{Potential, impedance, and rectification in membranes},
   J.\ Gen.\ Physiol.\ \textbf{27}, 37 (1943).

\bibitem{Arndt1970}
   R.A.\ Arndt, J.D.\ Bond and L.D.A.\ Roper,
   \textit{An exact constant-field solution for a simple membrane},
   Biophys.\ J.\ \textbf{10}, 1149 (1970).

\bibitem{Buck1976}
   R.P.\ Buck and S.\ Ciani,
   \textit{Electroanalytical Chemistry of Membranes},
   CRC Crit.\ Rev.\ Anal.\ Chem.\ \textbf{5}, 323 (1976).

\bibitem{Smith1977}
   J.R.\ Smith,
   \textit{Electrical characteristics of biological membranes in different
   environments},
   PhD thesis, University of New South Wales, 1977.

\bibitem{Brumleve1978}
   T.R.\ Brumleve and R.P.\ Buck,
   \textit{Numerical solution of the Nernst-Planck and Poisson equation system
   with applications to membrane electrochemistry and solid state physics},
   J.\ Electroanal.\ Chem.\ \textbf{90}, 1 (1978).

\bibitem{Eisenberg1999}
   R.S. Eisenberg,
   \textit{From Structure to Function in Open Ionic Channels},
   J.\ Membrane Biol.\ \textbf{171}, 1 (1999).

\bibitem{Samin2015}
   A.\ Samin and V.V.\ Subramaniam,
   \textit{Analytical Solutions to the Steady State Poisson-Nernst-Planck
   Equations in Electrobiochemical Systems},
   Appl.\ Phys.\ Res.\ \textbf{7}, 40 (2015).

\bibitem{Vetter1967}
   K.J.\ Vetter,
   \textit{Electrochemical Kinetics}
   (Academic, New York, 1967).

\bibitem{Bagotsky2006}
   V.S.\ Bagotsky,
   \textit{Fundamentals of electrochemistry}
   (Wiley, Hoboken, 2006).
   
\bibitem{Russel1989}
   W.B.\ Russel, D.A.\ Saville, and W.R.\ Schowalter,
   \textit{Colloidal Dispersions}
   (Cambridge University Press, Cambridge, 1989).

\bibitem{Hunter2001}
   R.J.\ Hunter,
   \textit{Foundations of Colloid Science}
   (Oxford University Press, Oxford, 2001).
   
\bibitem{Bazant2009a}
   M.Z.\ Bazant, M.S.\ Kilic, B.D.\ Storey and A.\ Ajdari,
   \textit{Towards an understanding of induced-charge electrokinetics at large
   applied voltages in concentrated solutions},
   Adv.\ Colloid Interface Sci.\ \textbf{152}, 48 (2009).

\bibitem{Mie1908}
   G.\ Mie,
   \textit{S\"{a}ttigungsstrom und Stromkurve einer schlecht leitenden
   Fl\"{u}ssigkeit},
   Ann.\ Physik \textbf{26}, 597 (1908).

\bibitem{Whitehead1930}
   J.B.\ Whitehead and R.H.\ Marvin
   \textit{The Conductivity of Insulating Oils},
   Trans.\ Am.\ Inst.\ El.\ Eng.\ \textbf{49}, 647 (1930).

\bibitem{Felici1985}
   N. Felici:
   \textit{High-field conduction in dielectric liquids revisited},
   IEEE Trans.\ Electr.\ Insul.\ \textbf{EI-20}, 233 (1985).

\bibitem{Lewis1994}
   T.J.\ Lewis,
   \textit{Basic Electrical Processes in Dielectric Liquids},
   IEEE Trans.\ Diel.\ Electr.\ Insul.\ \textbf{1}, 630 (1994).

\bibitem{Nernst1889}
   W.\ Nernst,
   \textit{Die elektromotorische Wirksamkeit der Jonen},
   Z.\ Physik Chemie \textbf{4}, 129 (1889).

\bibitem{Planck1890}
   M.\ Planck,
   \textit{Ueber die Erregung von Electricität und W\"{a}rme in Electrolyten},
   Ann.\ Physik Chemie \textbf{39}, 161 (1890).

\bibitem{Kornyshev2007}
   A.A.\ Kornyshev,
   \textit{Double-Layer in Ionic Liquids: Paradigm Change?},
   J.\ Phys.\ Chem.\ B \textbf{111}, 5545 (2007).

\bibitem{Kilic2007}
   M.S.\ Kilic, M.Z.\ Bazant and A.\ Ajdari,
   \textit{Steric effects in the dynamics of electrolytes at large applied
   voltages. I. Double-layer charging},
   Phys.\ Rev.\ E \textbf{75}, 021502 (2007).

\bibitem{Bazant2011}
   M.Z.\ Bazant, B.D.\ Storey and A.A.\ Kornyshev,
   \textit{Double Layer in Ionic Liquids: Overscreening versus Crowding},
   Phys.\ Rev.\ Lett.\ \textbf{106}, 046102 (2011).

\bibitem{Gillespie2002}
   D.\ Gillespie, W.\ Nonner and R.S.\ Eisenberg,
   \textit{Coupling Poisson–Nernst–Planck and density functional theory to
   calculate ion flux},
   J.\ Phys.\ Condens.\ Matter \textbf{14}, 12129 (2002).

\bibitem{Garvish2018}
   N.\ Gavish,
   \textit{Poisson-Nernst-Planck equations with steric effects - non-convexity
   and multiple stationary solutions},
   Physica D \textbf{368}, 50 (2018).

\bibitem{Suh2012}
   Y.K.\ Suh,
   \textit{Modeling and Simulation of Ion Transport in Dielectric Liquids --
   Fundamentals and Review},
   IEEE Trans.\ Diel.\ Electr.\ Insul.\ \textbf{19}, 831 (2012).


\bibitem{Yasufuku1979}
   S.\ Yasufuku, T.\ Umemura and T.\ Tanii,
   \textit{Electric conduction phenomena and carrier mobility bahavior in
   dielectric fluids},
   IEEE Trans.\ Electr.\ Insul.\ \textbf{EI-14}, 28 (1979).

\bibitem{Gafvert1992}
   U.\ Gafvert, A.\ Jaksts, C.\ Tornkvist and L.\ Walfridsson:
   \textit{Electrical Field Distribution in Transformer Oil},
   IEEE Trans.\ Electr.\ Insul.\ \textbf{EI-27}, 647 (1992).

\bibitem{Castellanos1998}
   A.\ Castellanos (ed.),
   \textit{Electrohydrodynamics}
   (Springer, Wien, 1998).
   
\bibitem{Butcher2006}
   M.\ Butcher, A.A.\ Neuber, M.D.\ Cevallos, J.C.\ Dickens and H.\ Krompholz:
   \textit{Conduction and Breakdown Mechanisms in Transformer Oil},
   IEEE Trans.\ Plasma Sci.\ \textbf{34}, 467 (2006).

\bibitem{Sha2014}
   Y.C.\ Sha, Y.X.\ Zhou, D.\ Nie, Z.\ Wu and J.G.\ Deng:
   \textit{A Study on Electric Conduction of Transformer Oil},
   IEEE Trans.\ Diel.\ Electr.\ Insul.\ \textbf{21}, 1061 (2014).

\bibitem{Gouy1909}
   M.\ Gouy,
   \textit{Sur la constitution de la charge \'{e}lectrique \`{a} la surface
   d'un \'{e}lectrolyte},
   C.R.\ Acad.\ Sci.\ \textbf{149}, 654 (1909).

\bibitem{Gouy1910}
   M.\ Gouy,
   \textit{Sur la constitution de la charge \'{e}lectrique \`{a} la surface
   d'un \'{e}lectrolyte},
   J.\ Physique \textbf{9}, 457 (1910).

\bibitem{Chapman1913}
   D.L.\ Chapman,
   \textit{A Contribution to the Theory of Electrocapillarity},
   Philos.\ Mag.\ \textbf{25}, 475 (1913).
   
\bibitem{Grahame1947}
   D.C.\ Grahame,
   \textit{The electrical double layer and the theory of electrocapillarity},
   Chem.\ Rev.\ \textbf{41}, 441 (1947).

\bibitem{Malvadkar1972}
   S.B.\ Malvadkar and M.D.\ Kostin,
   \textit{Solutions of the Nernst‐Planck Equations for Ionic Diffusion for
   Conditions near Equilibrium},
   J.\ Chem.\ Phys.\ \textbf{57}, 3263 (1972).

\bibitem{Buck1973}
   R.P.\ Buck,
   \textit{Steady-state space charge effects in symmetric cells with
   concentration polarized electrodes},
   Electroanal.\ Chem.\ Interf.\ Electrochem.\ \textbf{46}, 1 (1973).

\bibitem{Leuchtag1977}
   H.R.\ Leuchtag and J.C.\ Swihart,
   \textit{Steady-State Electrodiffusion},
   Biophys.\ J.\ \textbf{17}, 27 (1977).

\bibitem{Kosinska2008b}
   I.D.\ Kosi\'{n}ska, I.\ Goychuk, M.\ Kostur, G.\ Schmid and P.\ H\"{a}nggi,
   \textit{A singular perturbation approach to the steady-state 1D 
   Poisson-Nernst-Planck modeling},
   Acta Phys.\ Pol.\ B \textbf{39}, 1137 (2008).

\bibitem{Golovnev2009}
   A.\ Golovnev and S.\ Trimper,
   \textit{Exact solution of the Poisson–Nernst–Planck equations in the linear
   regime},
   J.\ Chem.\ Phys.\ \textbf{131}, 114903 (2009).

\bibitem{Kuzmin2010}
   R.N.\ Kuzmin, N.P.\ Savenkova, A.V.\ Shobukhov,
   \textit{Exact Steady States in the Electrodiffusive Model with Adsorptive
   Boundary Conditions},
   6th Int.\ Conf.\ Math.\ Modeling Comp.\ Sim.\ Mater.\ Technol., MMT-2010,
   23-27 Sep 2010, Ariel, Israel.

\bibitem{Golovnev2010}
   A.\ Golovnev and S.\ Trimper,
   \textit{Steady state solution of the Poisson–Nernst–Planck equations},
   Phys.\ Lett.\ A \textbf{374}, 2886 (2010).

\bibitem{Golovnev2011}
   A.\ Golovnev and S.\ Trimper
   \textit{Analytical solution of the Poisson–Nernst–Planck equations in the
   linear regime at an applied dc-voltage},
   J.\ Chem.\ Phys.\ \textbf{134}, 154902 (2011).

\bibitem{Shobukhov2014}
   A.V.\ Shobukhov and D.S.\ Maximov,
   \textit{Exact steady state solutions in symmetrical Nernst–Planck–Poisson
   electrodiffusive models},
   J.\ Math.\ Chem.\ \textbf{52}, 1338 (2014).

\bibitem{Wang2014}
   X.-Sh.\ Wang, D.\ He, J.J.\ Wylie and H.\ Huang,
   \textit{Singular perturbation solutions of steady-state
   Poisson-Nernst-Planck systems},
   Phys.\ Rev.\ E \textbf{89}, 022722 (2014).

\bibitem{Elad2019}
   D.\ Elad and N.\ Gavish,
   \textit{Finite domain effects in steady state solutions of
   Poisson-Nernst-Planck equations},
   SIAM J.\ Appl.\ Math.\ \textbf{79}, 1030 (2019).

\bibitem{Lyu2020}
   J.-H.\ Lyu, C.-C.\ Lee and T.-C.\ Lin,
   \textit{Near- and far-field expansions for stationary solutions of
    Poisson-Nernst-Planck equations},
   Math.\ Meth.\ Appl.\ Sci.\ \textbf{44}, 10837 (2020).

\bibitem{Asylamov2022}
   T.\ Aslyamov and M.\ Janssen,
   \textit{Analytical solution to the Poisson–Nernst–Planck equations for the
   charging of a long electrolyte-filled slit pore},
   Electrochim.\ Acta \textbf{424}, 140555 (2022). 

\bibitem{Jackson1999}
   J.D.\ Jackson,
   \textit{Classical Electrodynamics}
   (Wiley, New York, 1999).
   
\bibitem{Cohen1965}
   H.\ Cohen and J.W.\ Cooley,
   \textit{Time-Dependent Nernst-Planck Equations},
   Biophys.\ J.\ \textbf{5}, 145 (1965).

\bibitem{Debye1923}
   P.\ Debye and E. H\"{u}ckel,
   \textit{Zur Theorie der Elektrolyte},
   Phys.\ Z.\ \textbf{24}, 185 (1923).

\bibitem{McQuarrie2000}
   D.A.\ McQuarrie,
   \textit{Statistical mechanics}
   (Universal Science Books, Sausalito, 2000).

\bibitem{Gradshteyn1980}
   I.S.\ Gradshteyn and I.M.\ Ryzhik,
   \textit{Table of Integrals, Series, and Products}
   (Academic, New York, 1980).

\bibitem{Bocquet2002}
   L.\ Bocquet, E.\ Trizac and M.\ Aubouy,
   \textit{Effective charge saturation in colloidal suspensions},
   J.\ Chem.\ Phys.\ \textbf{117}, 8138 (2002).

\bibitem{Lide1998}
   D.R.\ Lide,
   \textit{Handbook of Chemistry and Physics, 79th ed.}
   (CRC, Boca Raton, 1998).

\bibitem{Stern1924}
   O.\ Stern,
   \textit{Zur Theorie der elektrolytischen Doppelschicht},
   Z.\ Elektrochemie \textbf{30}, 508 (1924).

\bibitem{Fedorov2008a}
   M.V.\ Fedorov and A.A.\ Kornyshev,
   \textit{Towards understanding the structure and capacitance of electrical
   double layer in ionic liquids},
   Electrochim.\ Acta \textbf{53}, 6835 (2008).

\bibitem{Fedorov2008b}
   M.V.\ Fedorov and A.A.\ Kornyshev,
   \textit{Ionic Liquid Near a Charged Wall: Structure and Capacitance of
   Electrical Double Layer},
   J.\ Phys.\ Chem.\ B \textbf{112}, 11868 (2008).

\end{thebibliography}
\end{document}